\documentclass{article}

\usepackage{graphicx}

\def\be{\begin{equation}}
\def\ee#1{\label{#1}\end{equation}}
\title{   3D gravity and non-linear cosmology }
\author{F. P. Devecchi\footnote{devecchi@fisica.ufpr.br},
 and  M. L.  Froehlich\\Departamento de F\'\i sica, Universidade Federal 
do Paran\'a\\
Caixa Postal 19044, 81531-990, Curitiba, Brazil}

\begin{document}

\maketitle

\begin {abstract} 

 By the inclusion of an additional term, 
non-linear in the scalar curvature $R$, it is  tested  if dark energy 
could rise
as a geometrical effect in 3D gravitational formulations.   
We  investigate  a cosmological fluid obeying a
non-polytropic
equation of state (the van der Waals equation) that is used to construct the 
energy-momentum tensor of the sources, representing the hypothetical inflaton 
in gravitational
interaction with a matter contribution.
 Following  the evolution in time of 
the scale
factor, its acceleration, and the energy densities of constituents 
it is  possible to construct  the description
 of an inflationary 3D universe, followed by a matter dominated era.
For later times it is verified that, under certain conditions, the 
non-linear term  in $R$ can generate 
the old 3D universe in accelerated expansion, where  
the ordinary matter is represented by the barotropic limit of the van der Waals constituent.

\end{abstract}

\noindent {040.50 +h; 98.80 -k}

\vskip 1cm

\maketitle

Theories of gravity in lower dimensions are useful instruments to investigate 
open questions in realistic theories\cite{Brown}. Looking from another perspective, they can
 be analysed as alternative theories
of gravity (and cosmology); in fact
several interesting (classical and quantum) results were obtained in 2D and 3D formulations 
\cite{Brown, Giddings, Cado}. 
In the particular  case of cosmological applications,
 the results include 
the description of a 3D Universe 
filled with ordinary  matter or/and electromagnetic
radiation\cite{Cornish, Deve6}, an inflationary 2D universe and a three-eras scenario, 
whose final 
accelerated 
period  could be 
associated to the effects of a dark energy constituent\cite{Deve2}.

 As is well-known, the investigation about the true nature of dark energy is a fundamental 
topic in 4D cosmology.
 One approach to describe this final accelerated era in 4D universes is to modify Einstein's 
theory of gravitation: the fundamental ingredient here is the inclusion of extra terms 
in the gravitational
dynamics that depend non-linearly in the curvature scalar $R$;
the cosmological
implications of this idea have been discussed in several works (for recent  investigations 
and   references see\cite{Vol}).
 These 
ideas come in substitution to the standard formulations that take the dark energy
as a ``usual'' source\cite{Vol}.
  This approach was 
tested in a 2D context
recently by these authors\cite{Deve5}, working with the Jackiw-Teitelboim (JT)
 model\cite{Brown, 
Jackiw}. 

If one wants to
follow this path in 3D models, starting with the 3D version of the Einstein-Hilbert theory,
 special care must be taken due to the presence some  peculiar features  
like the impossibility
 of gravitational propagation in free 
space and the absence of a Newtonian limit\cite{Brown, Giddings, Cornish}. On the other hand,
these problems can 
be solved   by taking 
as substitute the 3D scalar gravitational model\cite{Cornish, Deve6}. This theory offers 
several 
interesting 
results in 
cosmological applications like radiation-filled 3D universes and transition from inflationary
to matter dominated scenarios
\cite{Cornish, Deve6}.  Taking into account the arguments  above the proposal of our  work is 
to include a non-linear contribution in $R$ in the 
scalar 3D model,
and to investigate  in which cases  this non-linear term  is responsible 
for a final period of positive acceleration  playing the role 
of a dark energy constituent in 3D universes.

An important point when discussing 3D gravitational models is the existence 
of a  {Newtonian 
limit}\cite{Brown, Giddings, Cornish}. If we start with the 3D Einstein equations 
what is obtained (in the small velocities and weak field limits) 
is a {\it decoupling} of the gravitational field from sources
that leads  the  Newtonian potential to obey to the Laplace 
equation\cite{Giddings, Cornish}.

A Newtonian limit can be implemented in 3D gravity  if we substitute the 
Einstein 3D equations by the scalar model\cite{Brown, Cornish}, that was analised in a 
cosmological context in  
\cite{Cornish, Deve6}. A way of showing that a Newtonian limit indeed 
exists in 
this case  is to use the conformally-flat metric \cite{Brown}:

\begin{equation}
g_{\mu \nu}= \Phi^2(r)\eta_{\mu \nu} \approx 
[1+2\epsilon \theta(r)]\eta_{\mu \nu},
\end{equation}

\noindent where we invoke the weak field approximation, 
keeping terms up to 
first order in the parameter  $\epsilon$. The curvature scalar is in this case
\begin{equation}
R = -4\epsilon \frac{d^2}{dr^2}\theta (r)\, .
\end{equation}

\noindent In the weak field limit, we have that the trace of the energy-momentum tensor
of  sources ($T$) obeys
 $T\approx \rho$; so if it is required that in the Newtonian limit 
the gravitational
interaction be an atractive force, the scalar gravitational dynamics must be ruled by
\cite{Cornish}

\begin{equation}
R=-2\kappa T\, ,
\label{tror}
\end{equation}

\noindent 
 that leads to the well known non-relativistic expression  
$\nabla ^2 \theta =\kappa \rho$.
 Several cosmological solutions emerging from this scalar gravitational dynamics were 
analysed in \cite{Cornish,
Deve2, Deve6}. The results include (using the 3D Robertson-Walker metric) the 
description of periods dominated by 
matter and/or radiation\cite{Cornish}, inflationary young  3D universes and finally 
old 
universes dominated by dark energy in transition from a matter dominated 
period\cite{Deve6}; where 
the dark energy constituent is modeled using  different equations of state 
\cite{Cornish, Deve2, Deve6}.
  As we mentioned before the central idea of this work is to include in the field equations
a new term, non-linear  in the curvature scalar $R$. This discussion was done in 
\cite{Vol} for 4D models and in \cite{Deve5} for 2D models. We start with the new field 
equations

\begin{equation}
R^2 + 2\kappa TR - \omega ^2=0 \, ,
\end{equation}
where   $\kappa=2\pi G_2$ is the 3D gravitational coupling constant\cite{Giddings}
(with  $G_2=1$ in natural units). $\omega$ is a parameter that controls the non-linear term 
effects; in the 4D case some experimental data put restrictions on the values of this 
parameter, here it will be in principle free. 
Equation (4) has two roots in $R$; we mantain the one that preserves the
 limit
encoded in  equation (3).

 The usual hypothesis of 
homogeneity and isotropy,  in the form of the 3D Robertson-Walker metric,
$ds^2=dt^2-a^2(t)\left( \frac{dr^2}{1-kr^2}+r^2d\theta ^2\right)$, 
leave the gravitational field 
equations in the following form (for $k=0$)

\begin{equation}
2\frac{\ddot{a}}{a}+ \left(\frac{\dot{a}}{a}\right)^2=\pi T +
\frac{\sqrt{4\pi^2T^2+\omega^2}}{2}\, .
\end{equation}
\noindent   When we substitute the perfect-fluid energy-momentum-tensor trace  ($T = \rho - 
2p$) 
into the  equation (5) and into 
the  correspondent conservation law ($T^{\mu \nu};\nu =0$)
we obtain

\begin{equation}
\ddot{a}=-\frac{\dot{a}^2}{2a}-\frac{a\pi(\rho-2p)}{2}+
\frac{a\sqrt{4\pi^2(\rho-2p)^2+\omega^2}}{4}\, ,
\end{equation}

\begin{equation}
 \dot{\rho} + 2\dot a/a (p+\rho)=0\, .
\end{equation}

\noindent The early-times cosmic constituent is supposed to be ruled by the van der Waals 
equation of 
state\cite{Capo1},
$
p=\frac{b\rho}{1-\alpha \rho}
$, $b$ and $\alpha$ are constants, proposed in the cosmological context \cite{Capo1} to 
describe the behavior 
of a mixture of the  
inflaton with a matter constituent in a young universe. In fact,
 as the system evolves
the vdW equation approaches a barotropic equation of state (modeling  a period where
matter starts to predominate over the inflaton contribution)\cite{Capo1}.   

\begin{figure}[h]
\begin{center}
\includegraphics[width=10.3cm, height=7.7cm]{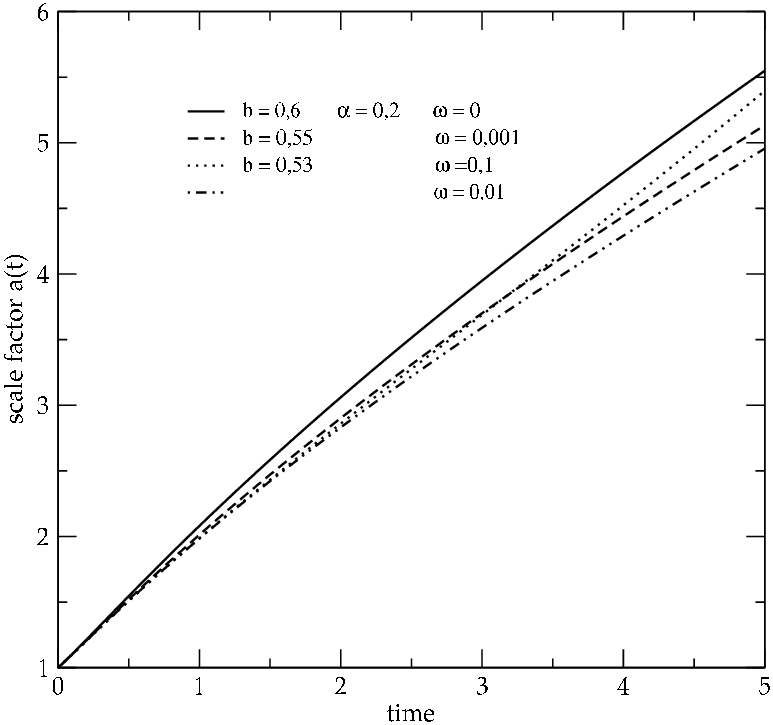}
\caption{Evolution in time of scale factor a(t)}
\end{center}
\end{figure}

\begin{figure}[h]
\begin{center}
\includegraphics[width=9.8cm, height=8.4cm]{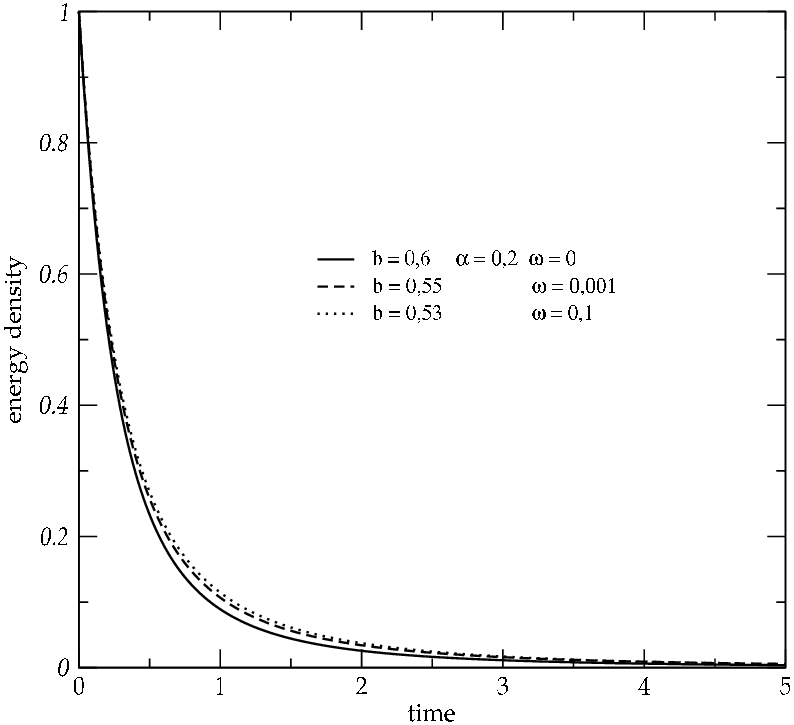}
\caption{Evolution in time of vdW energy density}
\end{center}
\end{figure}

The coupled system of differential equations (6) and (7),  is highly non-linear and we 
solve it numerically. 
The normalized boundary conditions are in this case 
: $a(0)=1$, $\dot{a}(0)=1$ e $\rho(0)=1$. These conditions simulate a young 3D universe, 
in the beginning
of an inflationary period\cite{Deve2, Deve5, Deve6}.

\begin{figure}[h]
\begin{center}
\includegraphics[width=10.6cm, height=8.5cm]{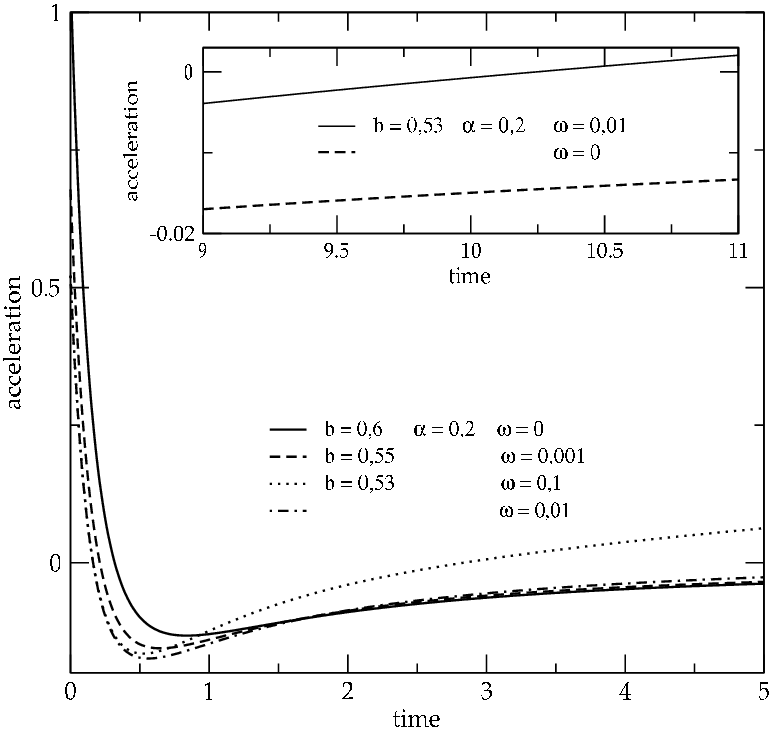}
\caption{Evolution in time of acceleration}
\end{center}
\end{figure}

 In figure (1) is represented the evolution in time of the scale factor $a(t)$ 
for several values of the
parameters.  What is observed  is that the increasing of the barotropic parameter $b$  
makes the 
expansion faster,
 a  result that is similar  to the one 
 found in the standard scalar case \cite{Deve6}; 
besides, the results  show that the increase of parameter  $\alpha$ also implies into a 
faster expansion (not displayed
 in the figure). First, we see
 that the modifications in the non-linear  term in $R$ (controlled through the 
parameter $\omega$) 
give no sensible results for early times ( in fact $1/R$ is negligible for early times)
 and therefore the 
standard 3D
 results\cite{Cornish, Deve6} are reproduced for a young 3D universe. On the other hand, 
for later times, the $\omega$ term promotes a faster expansion and, more importantly, a 
modification on the behavior of the acceleration, as we show bellow.

\noindent In figure (2) we show the behavior of the  energy density. In agreement with 
the 3D universe expansion, the
vdW energy density function is always decreasing. Accordingly, smaller values of  $b$ make
 the fall of the curve 
smoother, and bigger values of $\alpha$ make the fall more abrupt.

\vskip.5cm

Figure (3) shows the evolution in time of the acceleration. What is seen in the main 
picture is the 
first accelerated period and the first transition, 
from a 3D universe 
dominated by the inflaton to a period dominated by
matter (decelerated period). It is confirmed  that a second (decelerated) period exists, 
dominated by matter 
(as the vdW equation of state
  approaches 
a barotropic form). Till the end of this second period the non-linear  term in $R$ 
gives  a negligible 
contribution to the dynamics, following the results obtained in \cite{Deve6}. A different
and interesting situation is found for later times, as it is shown
in the small window: Increasing values of  $b$ imply 
into a 
slower fall of the acceleration for early times and a slower transtion to the
 third period for later times. More importantly, the link between dark energy and 
 a geometrical origin can be done. In fact, 
we have 
that the non-linear contribution in $R$ (controlled by the free parameter$\omega$)   
is responsible for the 
transition to the final  accelerated (third) period; increasing values of $\omega$ promote 
an earlier transition, 
making the
matter-dominated era shorter. As desired, when $\omega = 0$ there is no final transition 
at all. 
In summary,
 we have investigated, for 3D cosmological models,  the possibility of giving to 
the contributions
of dark energy  a geometrical origin, adding a non-linear term in the scalar curvature
 $R$ 
to the dynamics. We have investigated and confirmed the presence of  
transitions to accelerated
 periods, using the 3D scalar formulation with the addition of a particular  $1/R$ term, 
as a simulation 
of the beginning of 
a dark-energy dominated era; the
 modelling of a
 inflationary period was also possible. 
One constituent (ruled by the vdW equation) is enough to simulate the  3D
young universe   leaving the inflationary period. This is also the case to  describe the  
transition to a final 
period (leaving the matter
dominated era) when the non-linear 
term in $R$ starts to show its effects and promotes
the desired accelerated regime, that would correspond to the dark energy dominated era. 
 The results have also shown that  the role of the non-linear term is similar to the one 
found 
in 2D (when it is used the JT model \cite{Deve5}); although this parallel can be done only 
qualitatively:  different
forms in the explicit expressions for the dynamics are found
and also different   values of parameters are necessary to select the more interesting
scenarios. In both cases the analysis and 
results remain  as theoretical 
explorations of hypothetical universes in lower dimensions.


\newpage


\begin{thebibliography}{99}

\bibitem{Brown}  D. Grumiller et al., Phys. Rept. {\bf 369}, 327(2002); J.D. 
Brown, {\it Lower dimensional gravity}, World Scientific, Singapore, 1993.


\bibitem{Giddings}S. Giddings,J. Abbott and K.  Kuchar,
{Gen. Rel.  Grav.} {\bf 16}, 8 (1983); J.D. Barrow, A.B. Burd, 
D. Lancaster, Class. Quant. Grav. {\bf 3}, 551 (1986).

\bibitem{Cado} R. B. Mann and S. F. Ross, Phys. Rev. D {\bf 47}, 3312
(1993);  M. Cadoni and S. Mignemi,
Gen. Rel. Grav. {\bf 34},  2101 (2002).


\bibitem{Cornish} N.J. Cornish and N.E. Frankel, Phys. Rev.  {\bf D 43}, 2555
(1991); G.M. Kremer,  and F.P. Devecchi { Phys. Rev.}
{\bf D 65}, 12 (2002).

\bibitem{Deve2} M.H. Christmann, F.P. Devecchi, G.M.
Kremer, and C.M. Zanetti, Europhys. Lett. {\bf 67}, 728 (2004).




\bibitem{Vol} S. M. Carroll, V. Duvvuri, M. Trodden, M.S. Turner, Phys.Rev. D70 (2004) 043528;
 S. M. Carroll, A. De Felice, V. Duvvuri, D. Easson, M. Trodden, M. S. Turner,
Phys.Rev. D71 (2005) 063513; D. N. Vollick, Phys. Rev. D {\bf 68} 063510 (2003);
 D. S. Alves and G. M. Kremer, J. Cosmol. Astropart.
Phys. {\bf 10}, 009 (2004).

\bibitem{Deve5} F.P. Devecchi and M.L. Froehlich, Europhys. Lett. {\bf 71}, 179  (2005).

\bibitem{Deve6} F.P. Devecchi, G.M.
Kremer and C.M. Zanetti, "Accelerated regimes in 3D cosmologies", 
gr-qc/0509044, to appear in Gen. Relat. and Grav.




\bibitem{Jackiw} C. Teitelboim, in {\it Quantum Theory of Gravity}, edited by S. Christensen (Hilger, Bristol),p. 327, 
(1984); R. Jackiw, {\it ibid.} p. 403. 

\bibitem{Capo1} S. Capozziello, S. Martino, M. Falanga, Phys. Lett.  
{\bf A 299},494 (2002).





\end{thebibliography}
\end{document}